# Physical mechanisms of ohmic contact and tunnel diode: A novel explanation in terms of impurity-photovoltaic-effect resulting from infrared self-emission at room-temperature

Jianming Li

Institute of Semiconductors, Chinese Academy of Sciences, A35 Qinghua East Road, Haidian District, Beijing 100083, P. R. China

**ABSTRACT**

A mechanism of quantum-mechanical tunneling is based on electron-wavefunction and is used to explain ohmic contact as well as tunnel and Zener diodes. Tunneling is the important example of wave-particle duality. In this study, an attempt is made to explain these devices in particle description. As is well known, any object at room-temperature emits infrared (IR) photons due to blackbody radiation. The process of heavy doping can cause a lot of defects, e.g. vacancies and interstitials. The self-absorption of the IR emission could be achieved through sub-band-gap excitations due to defect-related levels in forbidden energy gap, creating carriers. In a heavily doped p-n junction diode, some of the IR-generated carriers diffuse into the junction which has a built-in field in a depletion layer. The built-in field then sweeps out the carriers, producing IR photocurrent. The IR photocurrent is regarded as the reverse current of the p-n junction according to the precedent where impurity-photovoltaic-effect is used to explain circuit devices (*arXiv*:2510.18226). Also, a reverse bias voltage increases depletion layer width, but carrier diffusion length out of the depletion layer remains unchanged, and more IR-generated carriers created in and near the depletion layer can contribute to the IR current. In addition, heavy doping results in not only a great many IR-generated carriers but also avalanche effect at low voltage. As a result, total reverse current dramatically increases with voltage, thus a Schottky diode can be altered to create an ohmic contact and a p-n diode can become a tunnel diode. Considering the wave-particle duality, the combination of quantum-mechanical and photovoltaic mechanisms is suggested to explain these devices. Each mechanism plays a big or small role in the explanation.

## Ⅰ. INTRODUCTION

Tunneling transmission is the important example of wave-particle duality [1]. Tunneling effect is used to explain some heavily doped semiconductor devices [2], such as ohmic contact, as well as tunnel diode, Zener diode, and resonant tunnel diode, etc. In quantum mechanics, tunnelling is attributed to an electron wavefunction that extends into a potential barrier. In the past, tunneling cannot be explained in particle description. In this study, an attempt is made to explain the semiconductor devices in particle description.

Metal-semiconductor contact has been studied nearly 100 years in order to achieve excellent ohmic contact. Ohmic contacts are indispensable for all semiconductor devices. A high resistance of metal-semiconductor contact is often a factor limiting the performance of semiconductor devices. As devices are miniaturized for advanced integrated circuits, the devices current density usually increases. This demands not only smaller ohmic resistance, but also contact made on a smaller area, which might present uniformity problem. The challenge for fabricating good ohmic contact has been increasing with device miniaturization. Various technologies have been developed for ohmic contacts [3-8]. More detailed experimental and theoretical studies are necessary to learn how to produce low-resistance, reliable, and reproducible contacts.

Tunnel diode was discovered by L. Esaki in 1958 [9], and is often called Esaki diode. Esaki explained the characteristics of Esaki diode by the quantum-mechanical tunneling concept [10]. Esaki diodes are used in microwave applications, etc.

Zener diode is one of the semiconductor diodes used in our daily life. Zener postulated a breakdown effect in a paper published in 1934 [11]. A Zener diode has a well-controlled breakdown voltage [12]. Thus, Zener diodes are used as voltage regulators.

## Ⅱ. SUPPOSED MECHANISM

There is already a precedent where impurity-photovoltaic-effect is used to explain circuit devices [13]. In this study, both ohmic contact and Esaki diode are explained in terms of impurity-photovoltaic-effect as follows.

As is well known, any object at room-temperature emits infrared (IR) photons due to blackbody radiation. Some research results [14,15] support that semiconductor devices at room temperature emit IR photons. It is also known that doping process can induce defects such as vacancies, interstitials, and imperfections associated with impurities, etc.

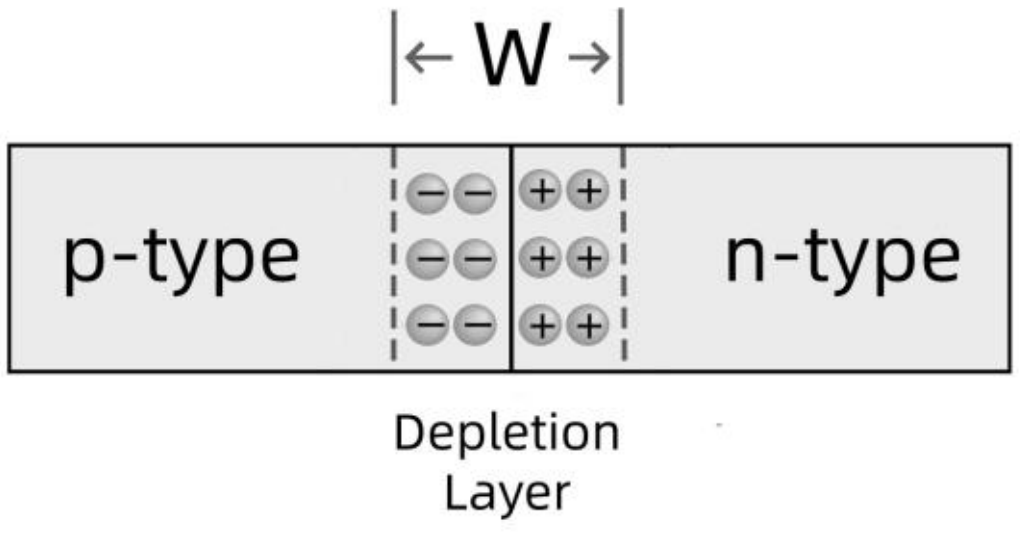

Fig. 1. A schematic of a p-n junction in thermodynamic equilibrium (without bias voltage), showing neutral regions and a depletion layer (a space-charge region). W is depletion layer width.

In a p-n junction diode shown in Fig. 1, the unavoidable defects create intermediate levels (bandgap states) in forbidden energy band-gap. The self-absorption of the IR emission within the diode could be achieved through sub-band-gap excitations shown in Fig. 2, which produces electron–hole pairs. The sub-band-gap excitation is analogous to climbing ladder — a child with limited strength reaches a certain height by climbing a ladder. Some of the IR-generated electrons and holes in the neutral regions diffuse into a depletion layer (a space-charge region with a built-in electric field). In the depletion layer, the electrons and holes drift in opposite directions due to the built-in field. Therefore, the p-n junction separates the IR-generated electrons and holes, resulting in a reverse current (IR current). This mechanism based on impurity-photovoltaic-effect has been discussed in detail [13,14].

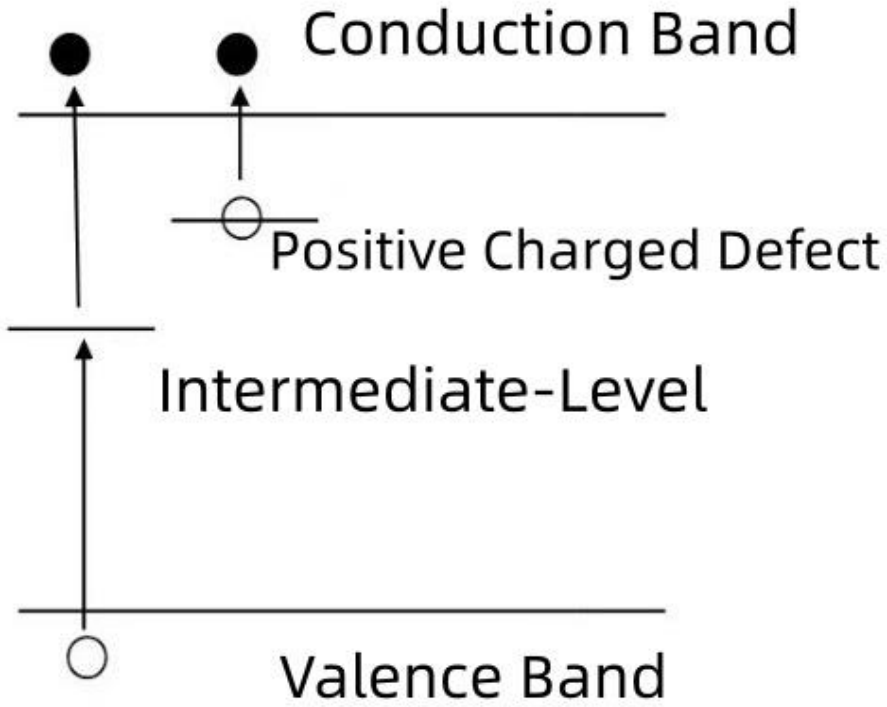

Fig. 2. The sub-band-gap excitation mechanism via the intermediate levels within forbidden gap.

## Ⅲ. DISCUSSION

As shown in Fig. 3, a depletion layer exists in a metal-semiconductor junction due to contact potential difference, resulting in a Schottky barrier diode which is electrically similar to the abrupt one-sided p-n junction diode. Here, a Schottky diode with heavy doping is discussed as follows.

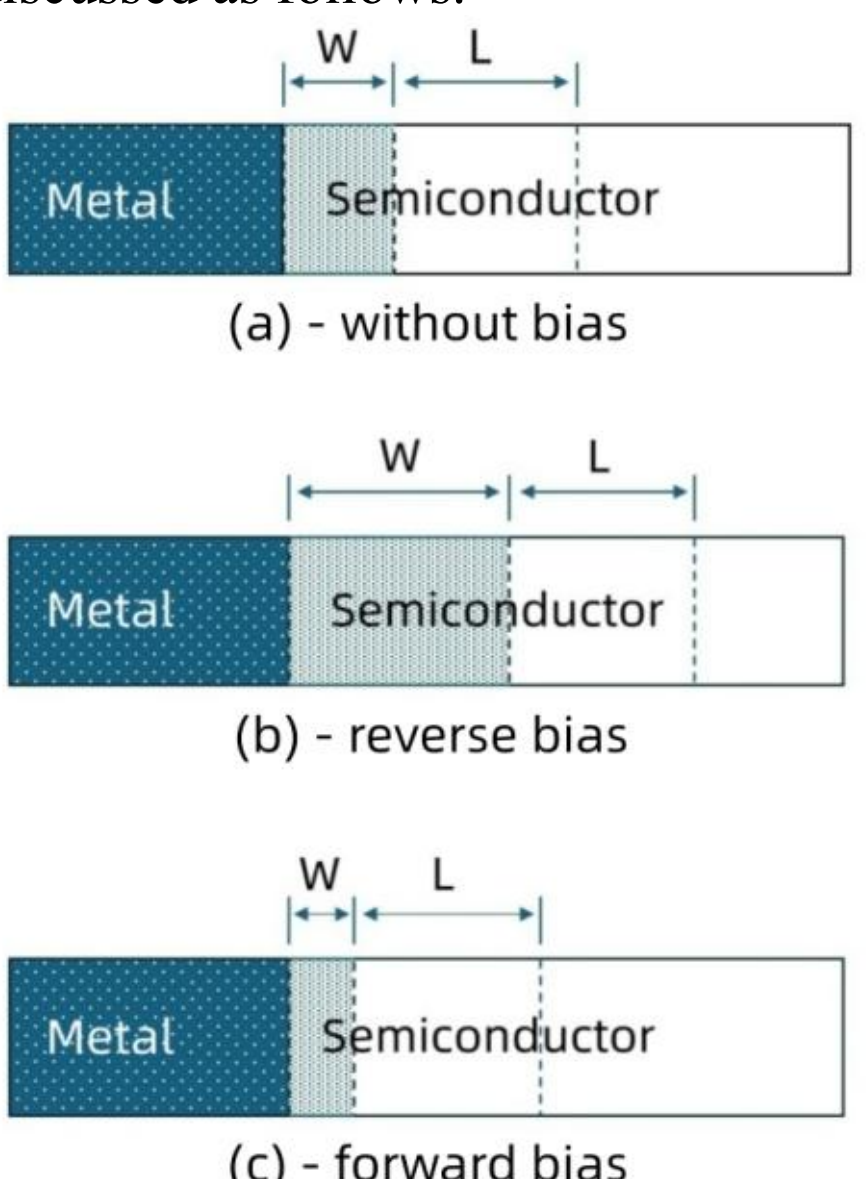

Fig. 3. A schematic of a metal-semiconductor junction. (a) Without bias, (b) reverse bias, and (c) forward bias. The shaded region in semiconductor is the depletion layer (space-charge region). W is depletion layer width. L is carrier diffusion length.

Under reverse bias, an applied voltage increases depletion layer width (W), but carrier diffusion length (L) on semiconductor side of the depletion layer remains unchanged (Fig. 3b). As a result, more IR-generated carriers created in and near the depletion layer can contribute to the IR current. On the other hand, heavy doping itself creates a lot of defects, contributing to the creation of a great many IR-generated carriers. If the doping is sufficiently heavy, the reverse drift current significantly increases with voltage. In addition, heavy doping results in a high gradient of charge concentration in the depletion layer with a strong built-in field, which causes avalanche effect at low voltage. In general, total reverse current dramatically increases with voltage.

In forward direction, the applied voltage decreases both W and the built-in field (Fig. 3c). Due to a high gradient of carrier concentration, a forward diffuse current of the carriers significantly surpasses the reverse drift current of the IR-generated carriers. Therefore, total forward current dramatically increases with voltage.

The metal-semiconductor junction with heavy doping has a small contact resistance for both forward and reverse directions, resulting in a non-rectifying current-voltage (I-V) characteristic shown in Fig. 4. A Schottky diode can be altered to create an ohmic contact, which is done through heavy doping. Furthermore, impurity-photovoltaic-effect applies to the explanation for the ohmic contacts made using all kinds of semiconductor materials.

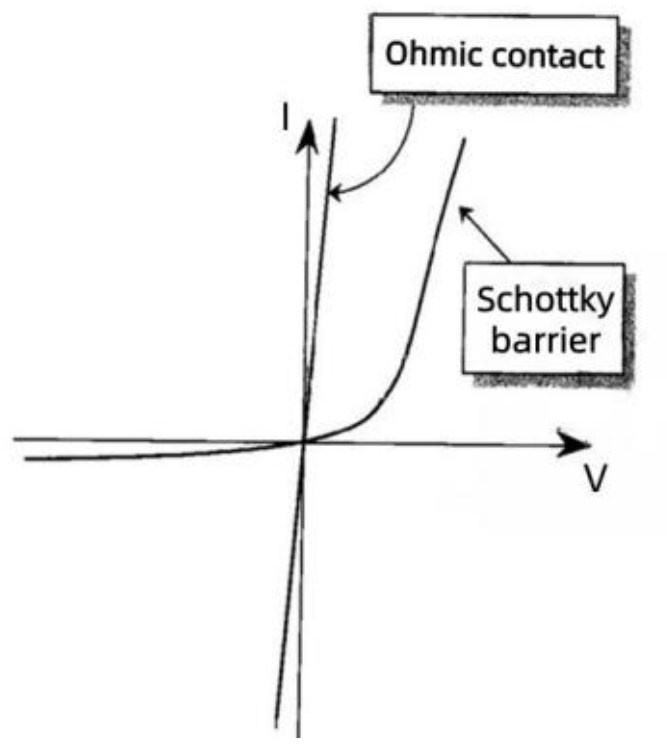

Fig. 4. I–V characteristics of an ohmic contact and a Schottky barrier diode.

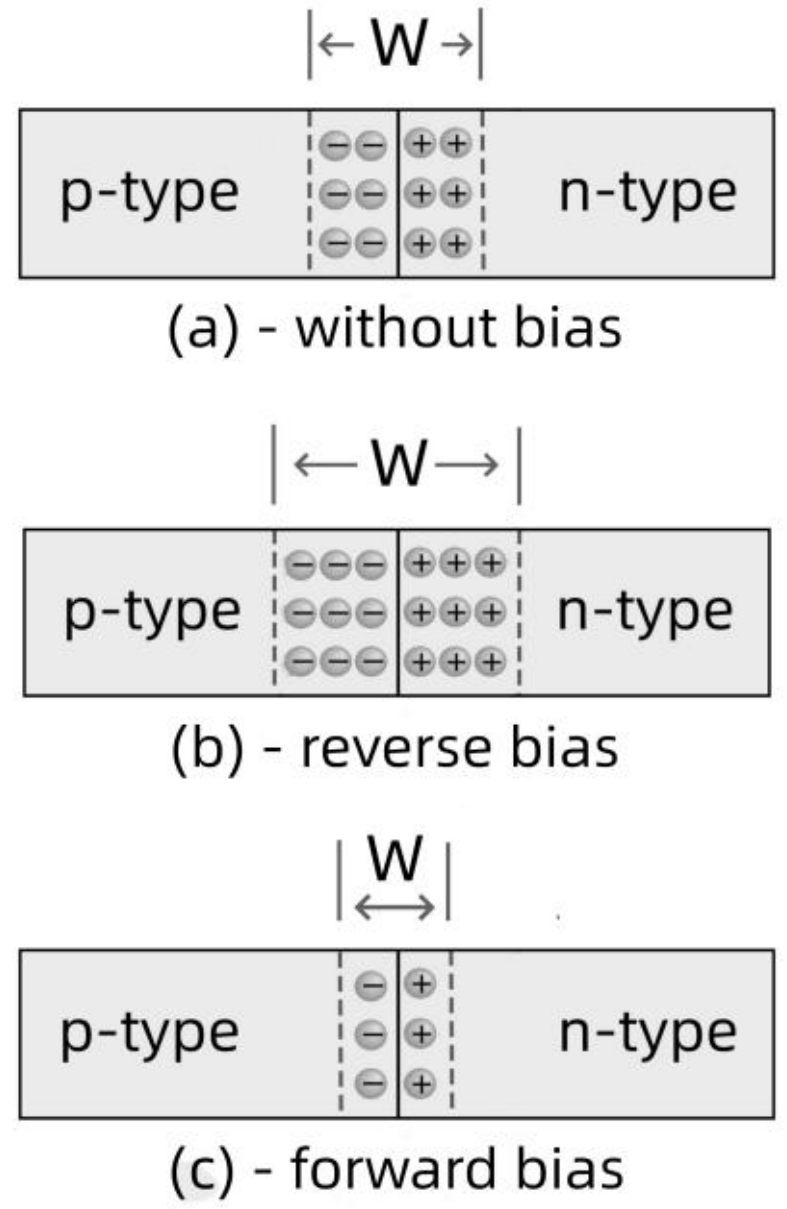

Fig. 5. A schematic of a p-n junction. (a) In thermodynamic equilibrium (without bias), (b) reverse bias, and (c) forward bias. W is depletion layer width.

Now, let us consider a p-n junction diode in which both p and n sides are heavily doped, as shown in Fig. 5. If both p and n regions of the p-n junction diode are so heavily doped that the I-V characteristic changes dramatically, the diode has a region of a negative differential resistance and thus is an Esaki diode. The I-V characteristic of Esaki diode is the result of two current components: forward diffuse current and reverse drift current. Here, Esaki diode is explained in terms of impurity-photovoltaic-effect as follows.

Under reverse bias, an applied voltage increases W (Fig. 5b), but carrier diffusion length on either side of the depletion layer remains unchanged. Therefore, more IR-generated carriers created in and near the depletion layer can contribute to the IR current. As in the ohmic contact discussed above, the reverse current dramatically increases with voltage (Fig. 6a).

In forward direction, the applied voltage decreases both W and the built-in field (Fig. 5c). Heavy doping results in a high gradient of carrier concentration in the junction, making a strong forward diffuse current. Such heavy doping causes a lot of defects, leading to the creation of many IR-generated carriers in the diode. However, numerous defects result in a high recombination rate. The quantity of the defects is so large that some of IR-generated carriers cannot finish the drift across the depletion layer due to defect-assisted recombination of IR-generated electrons and holes.

Upon application of a small forward voltage, W shrinks a little, and the forward diffuse current obviously surpasses the reverse drift current of IR-generated carriers.

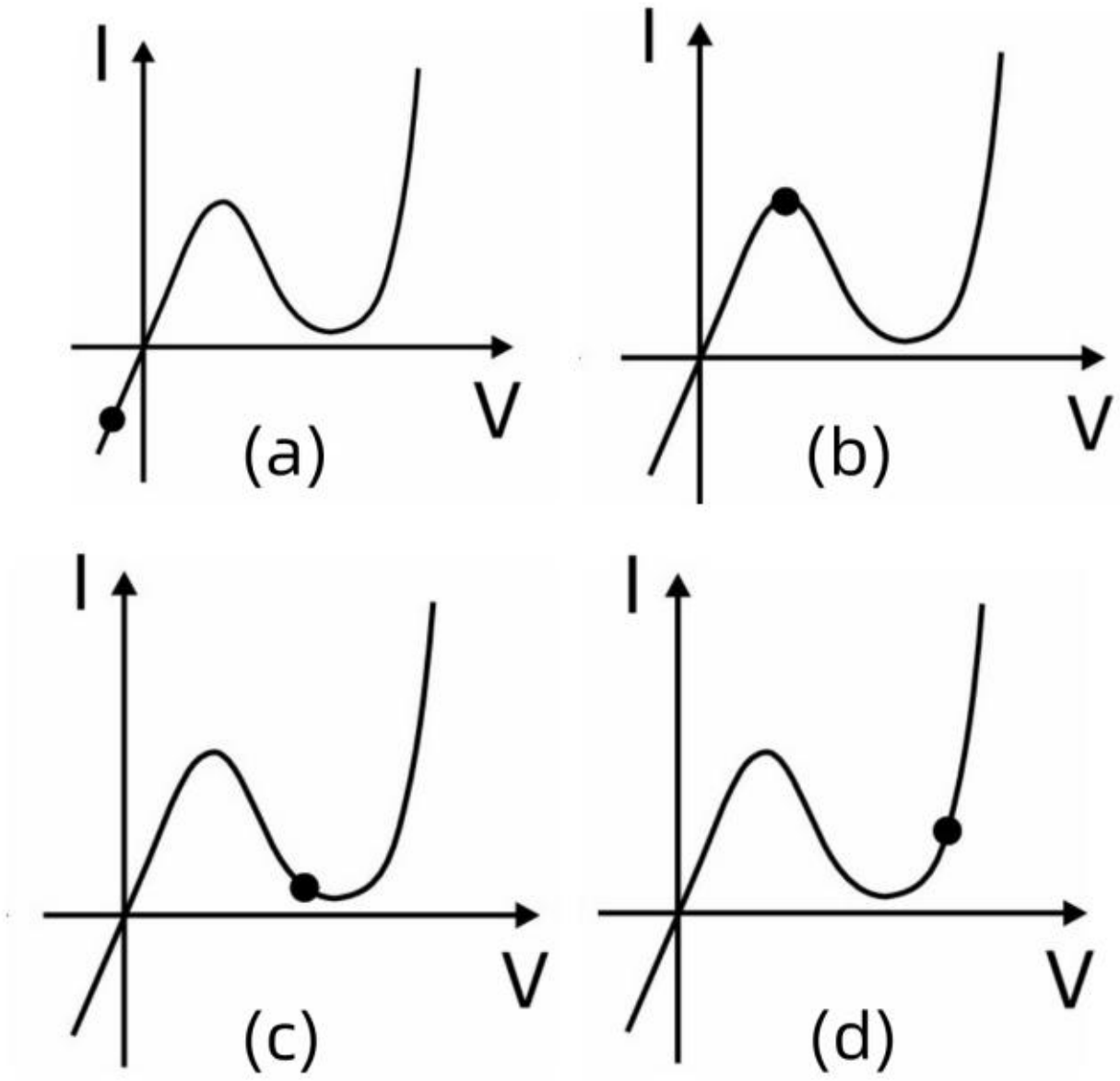

Fig. 6. I–V characteristics of an Esaki diode at various bias voltages as indicated with a dot in the I–V plot. (a) Reverse bias, (b) in the maximum of the current peak, (c) close to the current valley, and (d) forward bias with dominating diffuse current.

As the forward voltage increases, W is so narrow that more IR-generated carriers achieve the drift across the depletion layer for contributing the reverse current. Thus, some of the forward diffuse current can be counteracted by the reverse drift current, and net current will reach a maximum of peak current (Fig. 6b).

When the forward voltage is further increased, the net current decreases (Fig. 6c). If the forward voltage is applied such that the reverse drift current obtains the largest value, the I-V characteristic of the diode exhibits a minimum.

With still further increases of the applied voltage, the net current will increase exponentially with voltage and normal diode behaviour is regained (Fig. 6d).

In addition, other semiconductor devices, such as Zener diode, can also be explained in terms of impurity-photovoltaic-effect.

Comparing an ordinary diode, a Zener diode has higher doping concentration, leading to more IR-generated carriers and higher impact ionization rate. Thus, the impurity-photovoltaic-effect in the Zener diode enables avalanche effect to occur at lower voltage, creating Zener breakdown.

As reported in literatures [16-18], the tunneling current of Esaki diode has a weak temperature dependence. This can be explained on the basis of IR-emission related mechanism as follows.

The mobility of electrons in semiconductors is limited by lattice vibration scattering. An increase in temperature leads to stronger lattice vibrations. It is known that phonon scattering leads to an increasing resistivity with increasing temperature. Esaki diode has heavy doping in both p and n sides. The heavy doping results in numerous impurity-related defects. Defect and electron scattering processes can also increase resistivity with increasing temperature. These scattering processes can theoretically raise recombination rate of electrons and holes.

Let us first consider Esaki diode under reverse bias. Although the number of IR-generated carriers increases with increasing temperature, the change in reverse current with temperature may not be significant due to the effects of scattering and recombination.

For Esaki diode under forward bias, not only forward diffuse current but also reverse IR-current increases when temperature rises. The temperature-induced changes in the two currents can largely offset each other. Thus, net forward current may not change significantly with temperature.

Accordingly, it is theoretically possible that the current-voltage characteristic of Esaki diode has a weak temperature dependence.

## Ⅳ. CONCLUSIONS

In general, both ohmic contact and Esaki diode can be indeed explained by using the mechanism based on impurity-photovoltaic-effect in particle description. This study provides a reasonable support for the mechanism of impurity-photovoltaic-effect due to IR absorption. But quantum mechanics is a great discipline and provides an effective description of electrons in view of wave. Considering the wave-particle duality, the combination of quantum-mechanical tunneling and impurity-photovoltaic-effect mechanisms is suggested to explain the semiconductor devices described above. Each of the two mechanisms plays a big or small role in the explanation.

## DECLARATION OF COMPETING INTEREST

This is to certify that this manuscript does not make any conflict of interest with any person or institution or laboratories or any work.

## ACKNOWLEDGEMENT

This work was financially supported by Chinese Academy of Sciences (CAS).

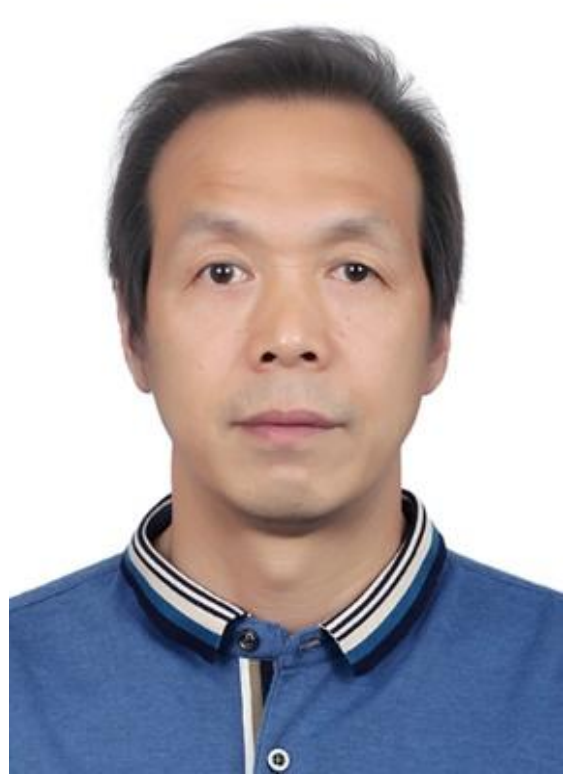

**Jianming Li**, research associate and senior engineer at the Institute of Semiconductors, Chinese Academy of Sciences. Email: jml_iscas@163.com; Tel.: +86-18310559676

Jianming Li was born in Beijing, China, in 1961. He studied at Beijing Normal University from 1980 to 1987. After graduated from the University, he joined the Institute of Semiconductors, Chinese Academy of Sciences. Since 1987, he has been engaged in research on photovoltaics, semiconductor materials and devices physics, etc. As a Visiting Scientist, he worked at Brookhaven National Laboratory, USA, from 1992 to 1994. As a Visiting Scientist, he worked at Pennsylvania State University, USA in 1997.